\documentclass[fleqn,10pt,twocolumn]{wlscirep}
\usepackage[utf8]{inputenc}
\usepackage[T1]{fontenc}

\DeclareCaptionLabelSeparator{bar}{\ | }
\DeclareCaptionLabelFormat{figlabel}{Fig.~#2} % automatic number
\usepackage{dsfont}
\usepackage{hyperref}

\title{Learning a dynamic four-chamber shape model of the human heart for 95,695 UK Biobank participants}

\author[1,2,*]{Qiang Ma}
\author[2,3]{Qingjie Meng}
\author[1,2]{Yicheng Wu}
\author[4,5]{Shuo Wang}
\author[2,6]{Mengyun Qiao}
\author[7]{Steven Niederer}
\author[8]{Declan P. O'Regan}
\author[1,9,10]{Paul M. Matthews}
\author[1,2,5]{Wenjia Bai}
\affil[1]{Department of Brain Sciences, Imperial College London, London, UK}
\affil[2]{Department of Computing, Imperial College London, London, UK}
\affil[3]{School of Computer Science, University of Birmingham, Birmingham, UK}
\affil[4]{Digital Medical Research Center, School of Basic Medical Sciences, Fudan University, Shanghai, China}
\affil[5]{Data Science Institute, Imperial College London, London, UK}
\affil[6]{Department of Mechanical Engineering, University College London, London, UK}
\affil[7]{National Heart and Lung Institute, Imperial College London, London, UK}
\affil[8]{MRC Laboratory of Medical Sciences, Imperial College London, London, UK}
\affil[9]{UK Dementia Research Institute, Imperial College London, London, UK}
\affil[10]{Rosalind Franklin Institute, Harwell Science and Innovation Campus, Didcot, UK}

\affil[*]{q.ma20@imperial.ac.uk}

\begin{abstract}
The human heart is a sophisticated system composed of four cardiac chambers with distinct shapes, which function in a coordinated manner. Existing shape models of the heart mainly focus on the ventricular chambers and they are derived from relatively small datasets. Here, we present a spatio-temporal (3D+t) statistical shape model of all four cardiac chambers, learnt from a large population of nearly 100,000 participants from the UK Biobank. A deep learning-based pipeline is developed to reconstruct 3D+t four-chamber meshes from the cardiac magnetic resonance images of the UK Biobank imaging population. Based on the reconstructed meshes, a 3D+t statistical shape model is learnt to characterise the shape variations and motion patterns of the four cardiac chambers. We reveal the associations of the four-chamber shape model with demographics, anthropometrics, cardiovascular risk factors, and cardiac diseases. Compared to conventional image-derived phenotypes, we validate that the four-chamber shape-derived phenotypes significantly enhance the performance in downstream tasks, including cardiovascular disease classification and heart age prediction. Furthermore, we demonstrate the effectiveness of shape-derived phenotypes in novel applications such as heart shape retrieval and heart re-identification from longitudinal data. To facilitate future research, we will release the learning-based mesh reconstruction pipeline, the four-chamber cardiac shape model, and return all derived four-chamber meshes to the UK Biobank.
\end{abstract}

\begin{document}

\flushbottom
\maketitle

\thispagestyle{empty}

\noindent 
Understanding the variations of cardiac anatomy and function, as well as their associations with disease risks and clinical outcomes remains one of the central challenges in cardiovascular research. The availability of large-scale cardiac imaging datasets in recent years, such as the UK Biobank Imaging Study \cite{collins2012ukbb,sudlow2015ukbb,bycroft2018ukbb,littlejohns2019ukbb}, provides an unprecedented opportunity to perform computational modelling of the cardiac shape and function at a population scale. Owing to excellent soft tissue contrast and non-ionising radiation, cine cardiac magnetic resonance (CMR) imaging employed by the UK Biobank study enables non-invasive and dynamic imaging to capture the cardiac shape and motion \cite{pennell2010cmr,petersen2016ukbb,kramer2020cmr}. Phenotypes derived from CMR images, such as the chamber volume, ejection fraction, and myocardial wall thickness, provide global quantitative description of the cardiac structure and function, serving as important biomarkers for cardiovascular research \cite{bai2018automated,bai2020population}. 
However, these phenotypes may overlook the detailed spatio-temporal (3D+t) characterisation of the cardiac shape and motion, which has been demonstrated to capture richer information compared to the global description of image-derived phenotypes \cite{young2009computational, puyol2017multimodal,bello2019motion,duchateau2020machine}. This motivates the cardiac imaging community to develop computational models to describe the 3D+t shape and motion of the heart \cite{xia2022msci,meng2023deepmesh,burns2024genetic,sorensen2024spatio,galazis2025atria,qiao2025meshheart}, which can facilitate a range of clinical and research tasks, including disease risk prediction \cite{bello2019motion}, clinical intervention planning \cite{prakosa2018personalized,boyle2019computationally}, biomechanical analysis \cite{nordsletten2011physics}, haemodynamic simulation \cite{taylor2013computational,kong2022mesh}, electrophysiological modelling \cite{trayanova2011whole,roney2022predicting,li2024solving}, cardiac digital twin modelling \cite{niederer2019computational,corral2020digital,niederer2021digital,qian2025digital}, etc.

Current computational shape modelling of the heart faces two major challenges. The first challenge is that existing cardiac shape models are mostly learnt from data at a relatively small scale. This may owe to both the lack of population-level cardiac imaging datasets and the absence of computational algorithms that can efficiently extract 3D shape from cardiac imaging data. Statistical shape models (SSMs) of the heart have been constructed using datasets ranging from dozens to thousands of subjects \cite{frangi2003ssm,lotjonen2004statistical,young2009computational,bai2015atlas,suinesiaputra2017ssm}. With the availability of UK Biobank population imaging data, cardiac shape modelling is currently being extended to tens of thousands participants \cite{xia2022msci,burns2024genetic,qiao2025meshheart}. In order to extract cardiac shapes from CMR images, guide-point model and atlas-based approaches have been explored in the past \cite{young2000left,bai2015atlas}. In recent years, a growing number of deep learning-based models have been developed for cardiac mesh reconstruction, which substantially improved the computational efficiency in processing large-scale datasets \cite{xia2022msci,meng2022mulvimotion,meng2023deepmesh,beetz2023multi,deng2023modusgraph,gaggion2025hybridvnet,qiao2025mesh4d}.

The second challenge is that existing cardiac shape models primarily focus on the ventricles of the heart \cite{meng2023deepmesh,burns2024genetic,qiao2025meshheart}, instead of covering all the four chambers including the left ventricle (LV), right ventricle (RV), left atrium (LA), and right atrium (RA). As one of the most widely used imaging modalities for cardiovascular research, cine CMR captures both cardiac shape and motion. However, the standard CMR imaging protocols typically do not provide a full coverage of the left and right atria \cite{pennell2010cmr,petersen2016ukbb,kramer2020cmr}. The two atria are captured only in a few long-axis view 2D image slices. The sparse coverage of the atria poses a technical challenge for estimating their 3D shapes. Although computed tomography (CT) imaging can cover all four cardiac chambers, it is cautiously used for only specific patient studies due to radiation burden \cite{hoogendoorn2012high,the2015scotheart,rodero2021linking,sorensen2024spatio}.  
To characterise both shape and motion of four cardiac chambers, existing work trained deep neural networks to learn four-chamber shape priors from CT datasets \cite{the2015scotheart} and completed sparse multi-view CMR segmentations into 3D dense four-chamber volumes \cite{xu2023complete,xu2024complete,muffoletto2024complete,ma2025cardiacflow}. Xia et al. built a four-chamber SSM by aligning a cardiac atlas to manually delineated contours and introduced MSCI-Net for four-chamber mesh reconstruction \cite{xia2022msci}. However, neither approaches enforced motion consistency across the cardiac cycle. Recent studies learnt 3D+t cardiac shape models from 4D CT \cite{sorensen2024spatio} or MR \cite{galazis2025atria} datasets. However, these studies were limited to the left atrium and trained on small-scale datasets only.

\begin{figure*}[!t]
\centering
\includegraphics[width=1.0\linewidth]{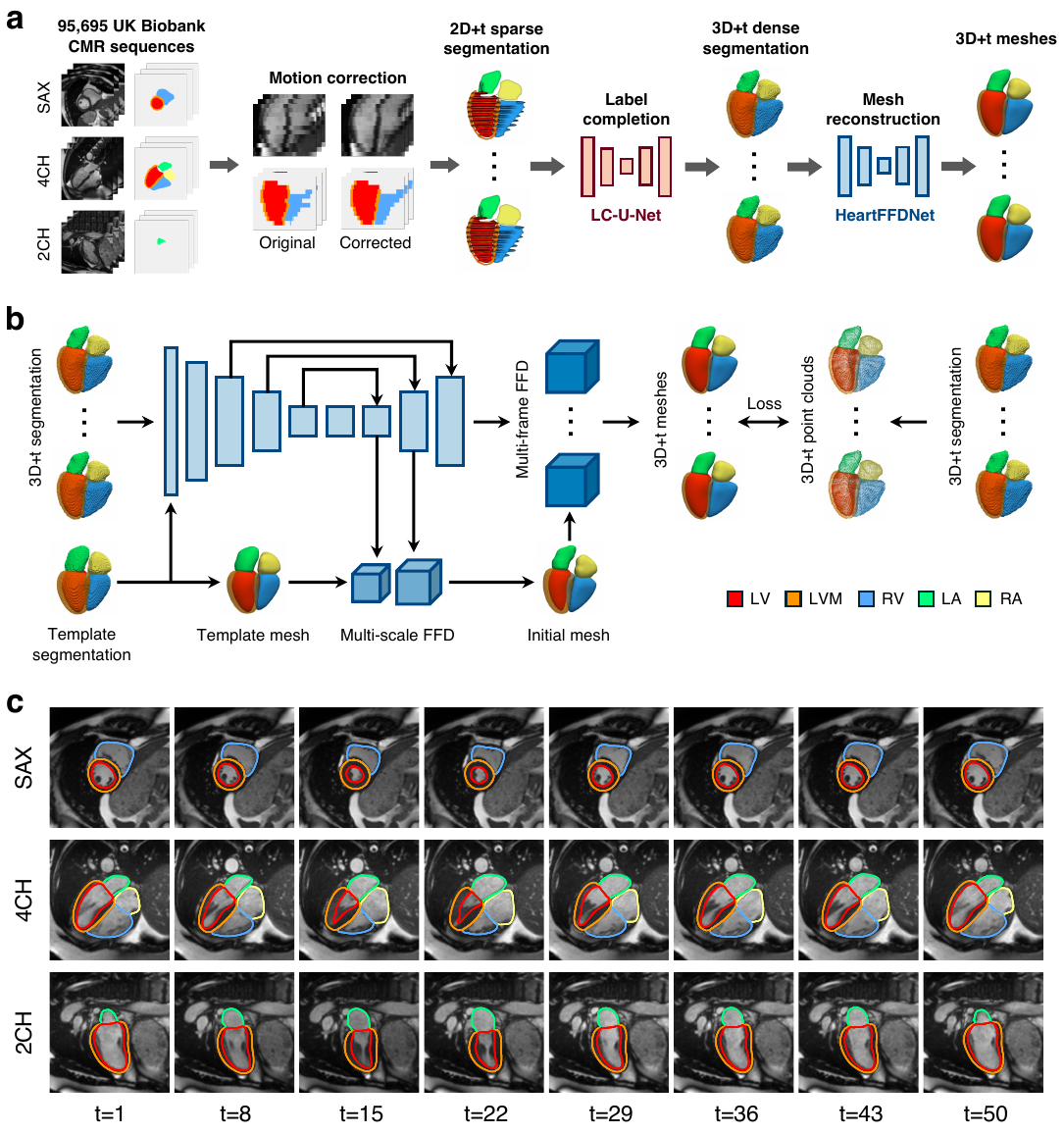}
\caption{\textbf{The deep learning-based pipeline for 3D+t cardiac four-chamber mesh reconstruction.} \textbf{a}, The processing pipeline consists of three components: motion correction for multi-view CMR images and segmentations, label completion from 2D+t multi-view sparse segmentations to 3D+t dense segmentations of the four chambers via a label completion U-Net (LC-U-Net), and finally 3D+t four-chamber mesh reconstruction with HeartFFDNet.
\textbf{b}, The architecture of HeartFFDNet. HeartFFDNet learns multi-scale and multi-frame free-form deformations (FFDs) from input 3D+t segmentations, warping a template mesh into four-chamber meshes across all time frames of a cardiac cycle. \textbf{c}, Qualitative visualisation of reconstructed 3D+t cardiac four-chamber meshes overlaid on multi-view CMR images. SAX: short-axis view; 4CH: long-axis four-chamber view; 2CH: long-axis two-chamber view.}
\label{fig:pipeline}
\end{figure*}

Here, we present \textit{HeartSSM}, a 3D+t statistical four-chamber shape model of the human heart learnt from a population-level CMR dataset from the UK Biobank Imaging Study with nearly 100,000 participants, the largest of its kind. To perform efficient population-scale shape modelling and address the sparse coverage problem of the atria, we develop a fast deep learning-based pipeline for four-chamber cardiac shape reconstruction from multi-view cine CMR images. The pipeline consists of several stages, including CMR motion correction, 3D cardiac label completion, and 3D+t mesh reconstruction. We employ the pipeline to reconstruct time-resolved four-chamber meshes for 95,695 UK Biobank participants after data cleaning, thereby curating a large-scale data resource for the shape modelling study. Based on the reconstructed meshes, we construct the HeartSSM shape model using incremental principle component analysis (PCA) to address the computational cost challenge \cite{weng2003ipca,zhao2006ipca,brand2006svd}.

We perform comprehensive evaluation of the deep learning analysis pipeline with respect to motion correction, label completion, and mesh reconstruction performance. For the HeartSSM shape model, we conduct correlation analyses of learnt principle components and vertex-wise shape variations with participant characteristics. We calculate \textit{shape-derived phenotypes} and compare them to conventional image-derived phenotypes \cite{bai2020population} in multiple downstream tasks, including cardiovascular disease classification \cite{qiao2025meshheart} and heart age prediction \cite{shah2023age}. Finally, we demonstrate the usability of the HeartSSM shape model in two novel tasks, namely heart shape retrieval and longitudinal heart re-identification. We envisage that the proposed deep learning-based mesh reconstruction pipeline, the curated 3D+t four-chamber shape dataset, and the developed cardiac shape model will provide a powerful foundation for future research in machine learning, 3D shape modelling, and cardiovascular studies.

\begin{figure*}[!t]
\centering
\includegraphics[width=1.0\linewidth]{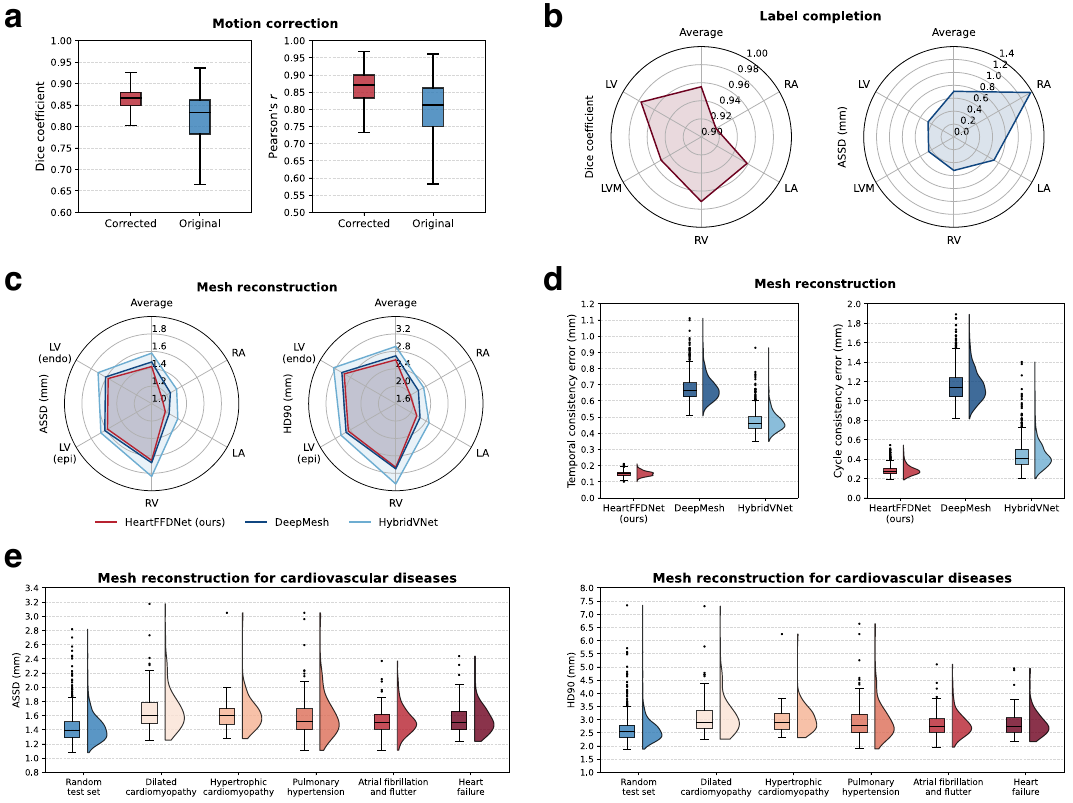}
\caption{\textbf{Performance of the 3D+t cardiac four-chamber mesh reconstruction pipeline.} \textbf{a}, Performance of motion correction measured by segmentation overlap (Dice) at the intersections of multi-view CMR segmentations and intensity consistency (Pearson's $r$) at the intersections of multi-view CMR images. \textbf{b}, Performance (Dice and ASSD) of 3D cardiac four-chamber label completion. \textbf{c}, Geometric errors (ASSD and HD90) of 3D+t four-chamber meshes of the proposed HeartFFDNet compared to DeepMesh and HybridVNet. \textbf{d}, Temporal consistency error and cycle consistency error of 3D+t four-chamber meshes reconstructed by HeartFFDNet compared to baseline methods. \textbf{e}, Geometric errors (ASSD and HD90) of the 3D+t meshes reconstructed by HeartFFDNet averaged across four cardiac chambers, evaluated on randomly selected test samples and participants with cardiovascular diseases.} 
\label{fig:eval}
\end{figure*}

\section*{Results}

\subsection*{Performance of 3D+t cardiac four-chamber mesh reconstruction pipeline}

We first provide an overview of the deep learning-based mesh reconstruction pipeline (Fig.~\ref{fig:pipeline}a), which takes CMR images and segmentations \cite{bai2018automated} as input and reconstructs 3D+t meshes of the four-chamber heart as output (see detailed description in Methods). The pipeline consists of three components: 1) CMR motion correction; 2) 3D four-chamber label completion; 3) 3D+t mesh reconstruction. First, an intensity-based motion correction algorithm \cite{villard2016correction} is employed and adapted to mitigate respiratory motion artefacts and multi-view misalignment in CMR images and segmentations. Then, a label completion U-Net (LC-U-Net) is applied to sparse multi-view CMR segmentations, including short-axis and long-axis views, to generate dense 3D+t segmentations of four chambers \cite{xu2023complete,xu2024complete,muffoletto2024complete,ma2025cardiacflow} (Supplementary Fig.~1). Finally, we develop a segmentation-to-mesh network, \textit{HeartFFDNet}, which takes 3D+t four-chamber segmentations as input, estimates multi-scale and multi-frame free-form deformations (FFDs), and warps a template mesh into motion consistent 3D+t meshes of four cardiac chambers (Fig.~\ref{fig:pipeline}b). HeartFFDNet provides a consistent mesh representation of four-chamber heart so that the predicted meshes share the same vertex connectivity across all time frames and across different subjects. The proposed deep learning pipeline is computationally efficient and only requires 2.87 seconds (s) to perform motion correction (0.32~s), label completion (2.4~s), and mesh reconstruction (0.15~s) for each CMR image sequence.

We evaluate the performance of each component of the cardiac four-chamber mesh reconstruction pipeline (Fig.~\ref{fig:pipeline}a). To evaluate the motion correction performance, we calculate the overlap of multi-view CMR segmentations at their intersection using the Dice coefficient, and the intensity consistency of multi-view CMR images at their intersection using the Pearson correlation coefficient ($r$). Fig.~\ref{fig:eval}a reports the Dice coefficient and the Pearson correlation coefficient before and after motion correction. For 95,695 CMR images with segmentations, the motion correction algorithm yields significantly improvement, with Dice increasing from $0.810\pm0.077$ to $0.859\pm 0.038$ (paired $t$-test, $p<10^{-307}$) and Pearson's $r$ increasing from $0.796\pm0.091$ to $0.861\pm 0.054$ (paired $t$-test, $p<10^{-307}$) after motion correction.

For 3D cardiac label completion, we trained a LC-U-Net \cite{xu2023complete,xu2024complete,muffoletto2024complete,ma2025cardiacflow} to complete the 3D cardiac four-chamber segmentation from 2D sparse multi-view segmentations. Since there are no ground-truth 3D segmentations for UK Biobank CMR images, we curate a dataset containing 3D four-chamber segmentations of CT images \cite{zeng2023imagecas,wasserthal2023totalsegmentator,Zhuang2019whs,Zhuang2019mvmm,tobon2015lasc,kiricsli2013casd,metz2009ccec} (see dataset information in Methods) for training and evaluation of the LC-U-Net. Fig.~\ref{fig:eval}b presents the Dice score and the absolute symmetric surface distance (ASSD) measured between the completed and ground-truth 3D four-chamber segmentations. The results show that the Dice scores of all four chambers are above 0.91, with the lowest performance observed for the RA (Dice, $0.918\pm0.028$; ASSD, $1.376\pm0.483mm$). The trained LC-U-Net is applied to the sparse 2D+t CMR segmentations from UK Biobank dataset to generate 3D+t dense four-chamber segmentations as illustrated in Fig.~\ref{fig:pipeline}a.

\begin{figure*}[!t]
\centering
\includegraphics[width=1.0\linewidth]{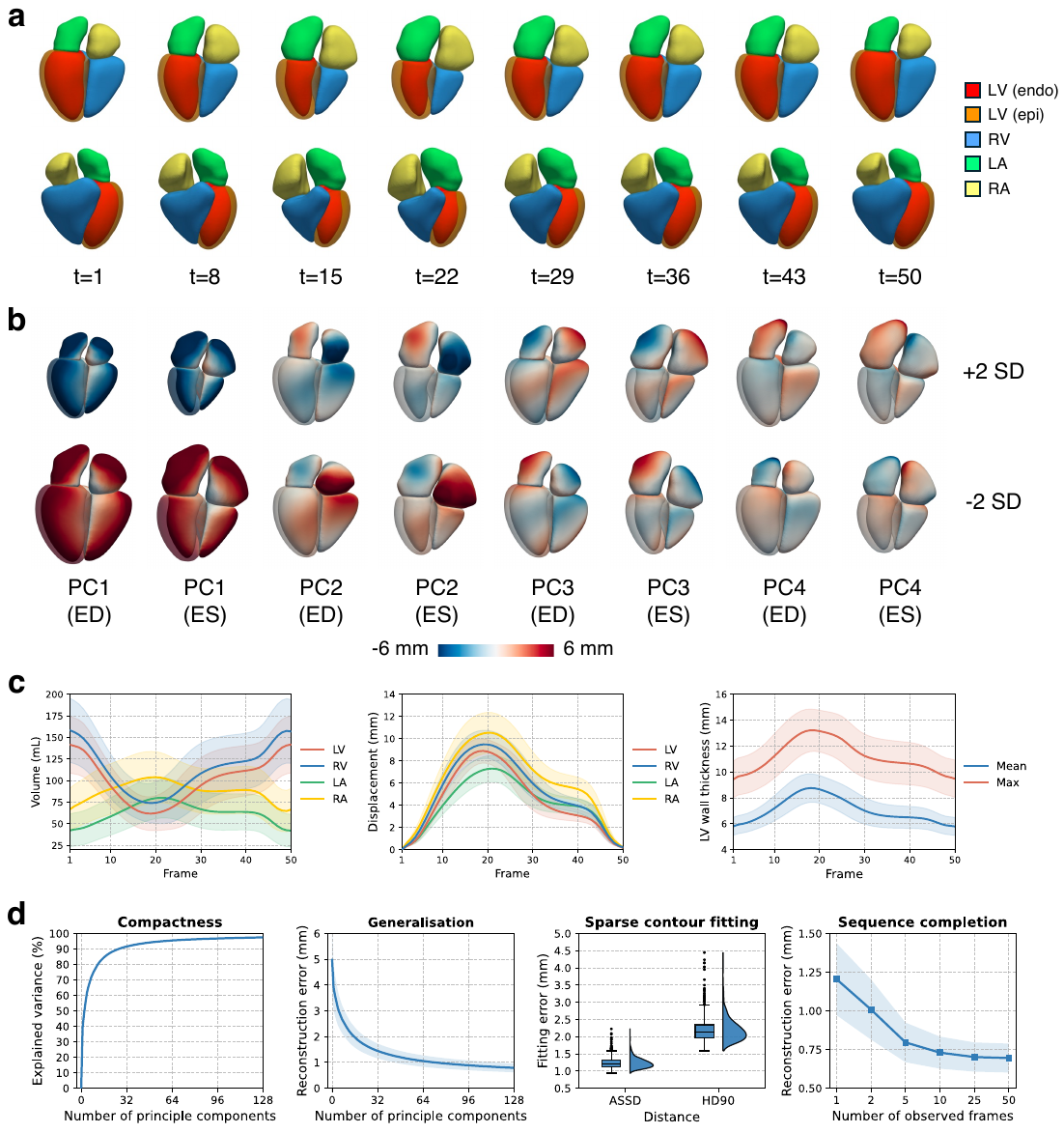}
\caption{\textbf{The 3D+t statistical shape model of the four-chamber heart.} \textbf{a}, The mean shapes (anterior and posterior view) of 3D+t four-chamber hearts derived from HeartSSM. 
\textbf{b}, The shape variations captured by the first four principle components (PC) of HeartSSM, with $\pm2$ standard deviation (SD) from the mean shape. The colour maps indicate the signed distances to the mean shape along normal directions. Red colour indicates inflation compared to the mean shape, while blue colour indicates contraction. Only end-diastolic (ED) and end-systolic (ES) frames are visualised.
\textbf{c}, The mean (line) and standard deviation (shaded area) of cardiac four-chamber volumes, displacements, and left ventricle wall thickness (mean and max) calculated from the HeartSSM fitted four-chamber meshes, displayed across all time frames and averaged over 95,695 UK Biobank participants. \textbf{d}, Performance of the HeartSSM. From left to right: the compactness of HeartSSM; the generalisation ability of HeartSSM on unseen cardiac shapes; performance of HeartSSM on 2D+t sparse contour fitting; performance of HeartSSM on sequence completion with partially observed frames. The shaded areas indicate $\pm1$ standard deviation.}
\label{fig:heartssm}
\end{figure*}

For 3D+t mesh reconstruction, we evaluate the geometric accuracy of the 3D+t four-chamber meshes predicted by HeartFFDNet. We extract 2D+t four-chamber contours from ground-truth UK Biobank CMR segmentations \cite{bai2018automated} with manual quality control and measure the distance between 2D+t contours and 3D+t meshes generated by HeartFFDNet, in terms of the uni-directional ASSD and the 90th percentile of Hausdorff distance (HD90). Fig.~\ref{fig:eval}c and Supplementary Table~3 compare the performance of HeartFFDNet to two recent baseline methods, DeepMesh \cite{meng2023deepmesh} and HybridVNet \cite{gaggion2025hybridvnet}, on the test set. The results show that HeartFFDNet achieves significantly better geometric accuracy (ASSD, $1.426\pm0.202mm$; HD90, $2.605\pm0.476mm$) than DeepMesh (paired $t$-test; ASSD, $1.478\pm0.249mm$, $p=3.5\times10^{-74}$; HD90, $2.697\pm0.561mm$, $p=2.2\times10^{-49}$) and HybridVNet (paired $t$-test; ASSD, $1.579\pm0.262mm$, $p=1.1\times10^{-242}$; HD90, $2.913\pm0.568mm$, $p=1.5\times10^{-211}$). An example of reconstructed 3D+t cardiac four-chamber meshes overlaid on multi-view CMR images is visualised in Fig.~\ref{fig:pipeline}c, showing that the reconstructed meshes align well with the anatomical boundaries of the four chambers on different CMR views.

In addition to geometric accuracy, we evaluate the temporal and cycle consistency of predicted 3D+t four-chamber meshes. The temporal consistency is measured by the temporal Laplacian error across all time frames of a cardiac cycle ($T=50$) to assess the coherence between consecutive frames. The cycle consistency is measured by the distance between the meshes at the starting frame and the end frame to assess whether the cardiac mesh returns to its original shape after a full cardiac cycle. Fig.~\ref{fig:eval}d demonstrates the temporal consistency error ($0.150\pm0.016mm$) and cycle consistency error ($0.284\pm0.052mm$) of HeartFFDNet, which outperforms both baseline methods (paired $t$-test, $p<10^{-183}$).

Furthermore, we examine the generalisability of HeartFFDNet for different subgroups within the UK Biobank populations with different cardiovascular diseases (determined using ICD-10 code), including the dilated cardiomyopathy (I42.0), hypertrophic cardiomyopathy (I42.1, I42.2), pulmonary hypertension (I27.0, I27.2), atrial fibrillation and flutter (I48), and heart failure (I50). Please see Supplementary Table~2 for detailed data information. The mesh reconstruction errors (ASSD and HD90) are reported in Fig.~\ref{fig:eval}e and Supplementary Table~4 for different disease groups. The mean geometric errors of the diseased cohorts are within $+1.3$ standard deviations (ASSD, $1.688mm$; HD90, $3.224mm$) of the geometric errors on a test set of randomly selected samples, demonstrating the generalisability of HeartFFDNet on participants with cardiovascular diseases.

\subsection*{3D+t statistical shape model of four cardiac chambers}
We construct a 3D+t SSM of the four cardiac chambers, named as \textit{HeartSSM}. Due to the high computational cost, HeartSSM is learnt by incremental PCA \cite{weng2003ipca,zhao2006ipca,brand2006svd}, using 64,000 3D+t four-chamber meshes generated by our learning-based pipeline as training data. HeartSSM captures the variations of both cardiac shape and motion in a large population, modelled by a mean 3D+t dynamic mesh sequence plus a linear combination of principle components (PCs). Fig.~\ref{fig:heartssm}a visualises the mean 3D+t heart shapes at different time frames, demonstrating the contraction and relaxation process of the four chambers across a cardiac cycle. Fig.~\ref{fig:heartssm}b visualises the first four PCs of HeartSSM with $\pm2$ standard deviation (SD) at the end-diastolic (ED) and end-systolic (ES) frames (see Supplementary Fig.~5-7 for additional PCs and time frames). The colour map indicates the vertex-wise signed distance of a shape compared to the mean shape, measured along the normal direction of each vertex. Red colour indicates inflation compared to the mean shape and blue colour indicates contraction. Fig.~\ref{fig:heartssm}b shows that PC1 controls the size of the whole heart, PC2 mainly affects the shape of the RA, and other PCs characterise different modes of shape variations. We employ HeartSSM to reconstruct 3D+t four-chamber meshes of 95,695 UK Biobank subjects and encode each of the 3D+t mesh sequence into a 128-dimensional shape descriptor. Based on the reconstructed meshes, we compute the four-chamber volumes, displacements, and LV wall thickness (mean and max) averaged over all subjects across a cardiac cycle (Fig.~\ref{fig:heartssm}c).

We assess the representation capacity of HeartSSM using four metrics \cite{davies2008ssm,booth2016face} in terms of compactness, generalisation, capacity for fitting sparse 2D+t contours and for completing 3D+t sequences with missing frames (Fig.~\ref{fig:heartssm}d). Regarding compactness, HeartSSM explains over 90\% of the variance with the first 27 PCs and 97.45\% of the variance with all 128 PCs. The generalisation ability is evaluated by fitting the HeartSSM to unseen test 3D+t cardiac meshes from the UK Biobank. We measure the reconstruction error by computing the mean squared error between the test 3D+t shapes and the shapes reconstructed by the HeartSSM. Fig.~\ref{fig:heartssm}d shows that the reconstruction error ($0.788\pm0.170mm$) is relatively low when using 128 PCs. 
We examine the fitting ability of HeartSSM using test subjects for sparse contour fitting, \emph{i.e.}, fitting 3D+t meshes to sparse 2D+t four-chamber contours, and for sequence completion, \emph{i.e.}, fitting 3D+t meshes to a mesh sequence with missing time frames. Qualitative illustrations of contour fitting and sequence completion are provided in Supplementary Fig.~3 and 4. By optimising the weights of the PCs, HeartSSM accurately fits onto 2D+t sparse contours with relatively low distance errors (Fig.~\ref{fig:heartssm}d; ASSD, $1.238\pm0.154mm$; HD90, $2.198\pm0.346mm$). For missing sequence completion, HeartSSM reconstructs 3D+t cardiac meshes effectively from even a single observed frame (Fig.~\ref{fig:heartssm}d; ASSD, $1.204\pm0.232mm$). More accurate fitting can be achieved with only 10 observed frames (Fig.~\ref{fig:heartssm}d; ASSD, $0.730\pm0.104mm$).

\begin{figure*}[!t]
\centering
\includegraphics[width=0.98\linewidth]{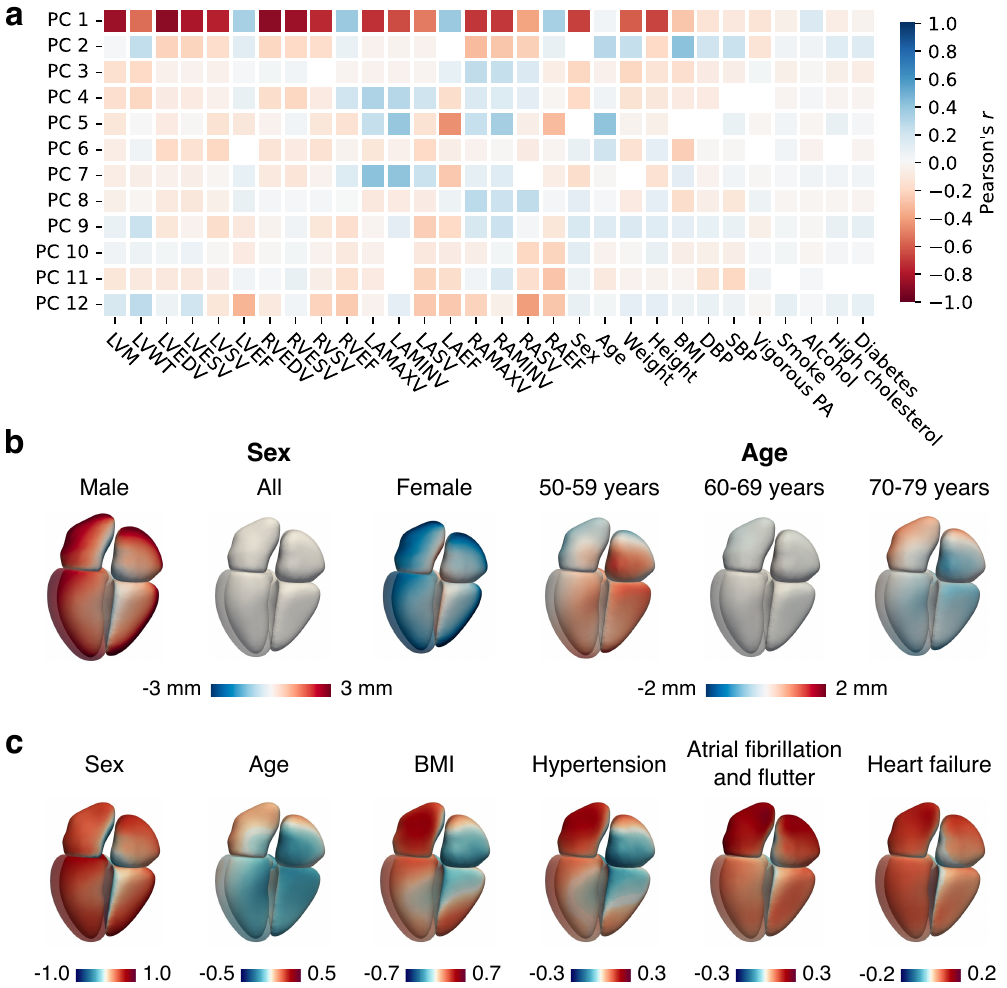}
\caption{\textbf{Correlation analysis of cardiac four-chamber shape model.} \textbf{a}, Correlations of PCs with four-chamber phenotypes, demographics, anthropometrics, and cardiovascular risk factors. Only significant correlations are visualised ($p<0.05$). \textbf{b}, Mean cardiac four-chamber shapes and shape variations averaged across all time frames for different sex and age groups. The colour maps depict the signed distances along surface normals to the mean cardiac shape of all participants. Red colour indicates inflation compared to the mean shape, while blue colour indicates contraction. \textbf{c}, Correlations of vertex-wise shape variations with sex, age, BMI, hypertension, atrial fibrillation and flutter, and heart failure. The colour maps visualise significant Pearson correlation coefficients after Bonferroni correction ($p<0.05/n$, $n=27,034$ denoting the number of all vertices). }
\label{fig:correlation}
\end{figure*}

\subsection*{Associations of four-chamber shapes with demographics, anthropometrics and diseases}
Using 3D+t four-chamber meshes reconstructed by HeartSSM, we calculate phenotypes characterising the heart structure and function, including the ventricular ED volumes (EDV) and ES volumes (ESV), atrial maximum volumes (MAXV) and minimum volumes (MINV), stroke volumes (SV), ejection fractions (EF), LV myocardial mass (LVM), as well as LV wall thickness (LVWT). We refer to these phenotypes as \textit{shape-derived phenotypes}, in contrast to the image-derived phenotypes estimated directly from CMR image segmentations \cite{bai2018automated,bai2020population}. 
To interpret the cardiac shape variations captured by the PCs, we explore the correlations between the first 12 PC weights (capturing $>80\%$ variance) of the HeartSSM with shape-derived phenotypes, sex, age, weight, height, body mass index (BMI), diastolic blood pressure (DBP), systolic blood pressure (SBP), vigorous physical activities (PA), smoke, alcohol, high cholesterol, and diabetes \cite{bai2020population,burns2024genetic}. The Pearson correlation coefficients are presented in Fig~\ref{fig:correlation}a. As expected, the first PC, which is negatively correlated to the size of the hearts (Fig.~\ref{fig:heartssm}b), shows negative correlation with LV mass, LV wall thickness, four-chamber volumes, sex, weight, and height. Positive correlations are observed between the first PC with four-chamber ejection fractions.

\begin{figure*}[!t]
\centering
\includegraphics[width=1.0\linewidth]{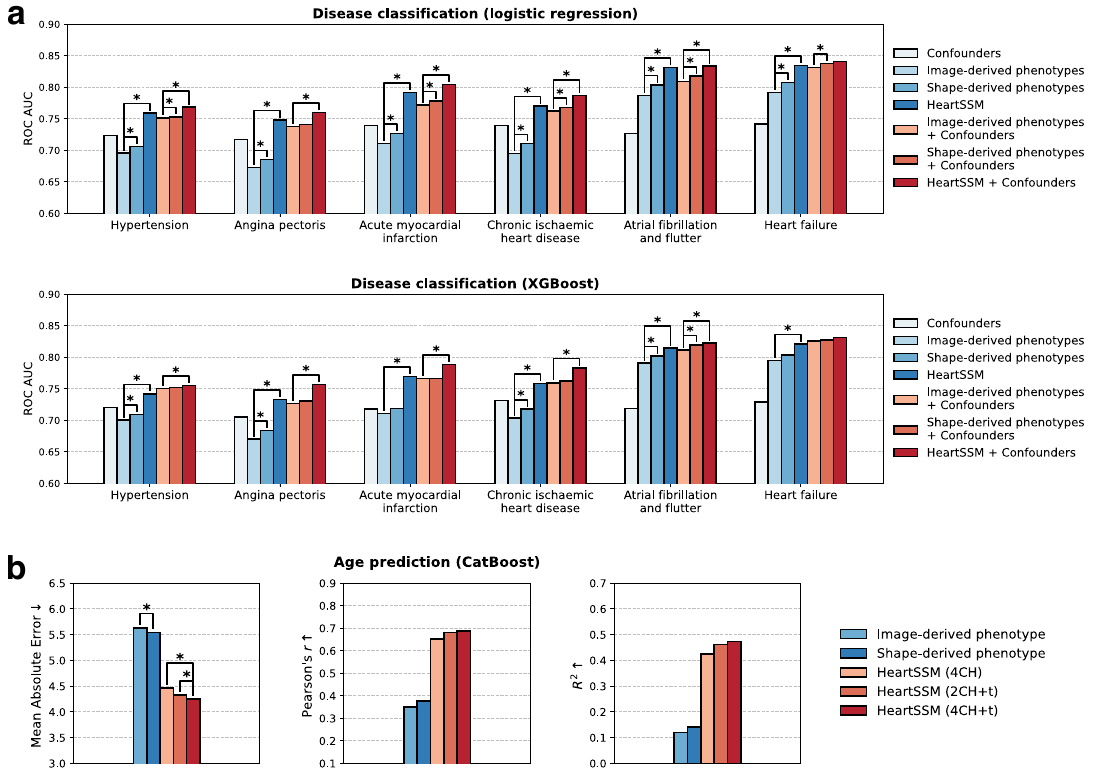}
\caption{\textbf{Performance for cardiovascular disease classification and chronological age prediction.} \textbf{a,} Binary classification performance for six cardiac diseases using different features, including image-derived phenotypes, shape-derived phenotypes, and HeartSSM shape descriptor, with or without confounders (sex, age, weight, height, and BMI). The ROC AUC scores for disease classification using logistic regression and XGBoost models are reported. The symbol * indicates a significant difference (two-sided DeLong test, $p<0.05$). \textbf{b,} Chronological age prediction performance using different features, including image-derived phenotypes, shape-derived phenotypes, 3D four-chamber (4CH) HeartSSM shape descriptor, 3D+t bi-ventricular (2CH+t) HeartSSM shape descriptor, and 3D+t four-chamber (4CH+t) HeartSSM shape descriptor. The mean absolute error, Pearson's $r$, and coefficient of determination $R^2$ for age prediction using CatBoost model are reported. The symbol * indicates statistically significant difference (paired $t$-test, $p<0.05$). }
\label{fig:classification}
\end{figure*}

We compare the shape variations between different sex and age groups in a cohort of 62,778 UK Biobank participants with available demographic and disease information. We compute the mean cardiac four-chamber shape for each group, averaged across all subjects of each group and across all $T=50$ time frames, as shown in Fig.~\ref{fig:correlation}b. The colour maps represent the shape variations compared to the mean cardiac shape of all subjects. Furthermore, we perform correlation analysis between the shape variation of each vertex and three clinical factors including sex, age, and BMI, as well as three disease types including hypertension (I10-I15), atrial fibrillation and flutter (I48), and heart failure (I50). At each vertex of the cardiac four-chamber meshes, we calculate the Pearson correlation coefficient ($r$) between its mean signed distance over all frames and each of the above six factors. The correlation maps are visualised in Fig.~\ref{fig:correlation}c.
Fig.~\ref{fig:correlation}b and c demonstrate that male hearts are generally larger than female hearts, and the ventricle volumes show a decreasing trend with increasing age. The correlation maps of shape variation with BMI and hypertension in Fig.~\ref{fig:correlation}c exhibit similar patterns across all four chambers. For atrial fibrillation and flutter, we observe a stronger correlation (two-sided $t$-test, $p<10^{-307}$) on the two atria ($r=0.108^\dagger$, $^\dagger$averaged across all vertices) compared to the two ventricles ($r=0.052^\dagger$). The positive correlation shows that the enlargement of the atria is positively associated with atrial fibrillation and flutter \cite{wozakowska2005af,gupta2014af}. There is a stronger positive correlation (two-sided $t$-test, $p<10^{-307}$) between the left heart shape variation with heart failure ($r=0.063^\dagger$) than the right heart shape variation ($r=0.019^\dagger$) \cite{friedberg2014failure,uk2016failure,bozkurt2021failure}.

\subsection*{Shape features for cardiovascular disease classification}
HeartSSM characterises each 3D+t four-chamber meshes using a 128-dimensional shape descriptor, \emph{i.e.}, the weights of the PCs. We investigate whether the shape derived-phenotypes and HeartSSM shape descriptor provide discriminative features for cardiovascular disease classification \cite{qiao2025meshheart}. We train two types of classification models, logistic regression and XGBoost \cite{chen2016xgboost} for the binary classification of six cardiovascular diseases including hypertension (I10-I15), angina pectoris (I20), acute myocardial infraction (I21), chronic ischaemic heart disease (I25), atrial fibrillation and flutter (I48), and heart failure (I50). The classifiers are trained with different features, including confounders (age, sex, weight, height, BMI), image-derived phenotypes, shape-derived phenotypes, and HeartSSM shape descriptor. Fig.~\ref{fig:classification}a and Supplementary Table~5 report the disease classification performance, evaluated using the area under the receiver operating characteristic curve (ROC AUC) in four-fold cross validation. The classification performance with different features are compared using the two-sided DeLong test \cite{sun2014delong,qiao2025meshheart}.
Across all six cardiac disease classification tasks, shape-derived phenotypes achieve consistently higher AUCs than image-derived phenotypes, regardless of the inclusion of confounders and the choice of classifiers. HeartSSM shape descriptor further improves classification performance compared to shaped-derived phenotypes. Combining both confounders and HeartSSM shape descriptors as features yields the best performance for classifying all six diseases.

\begin{figure*}[!t]
\centering
\includegraphics[width=1.0\linewidth]{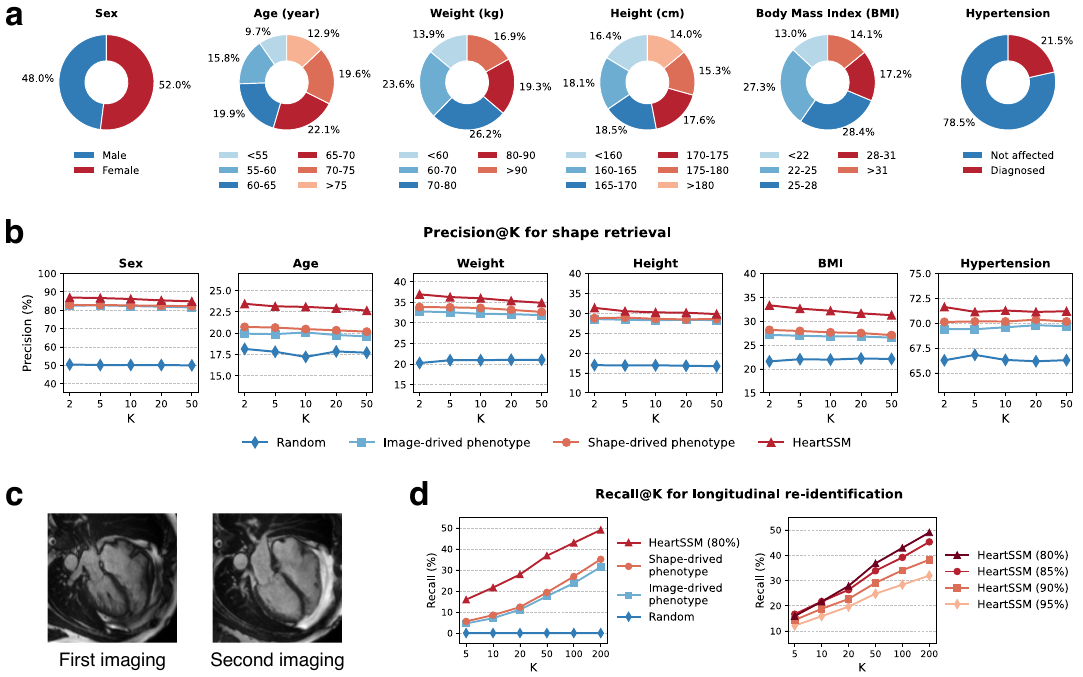}
\caption{\textbf{Performance on cardiac shape retrieval and longitudinal re-identification.} \textbf{a}, Sex, age, weight, height, BMI and hypertension are categorised into different groups. \textbf{b}, Top-$K$ precision for shape retrieval using random retrieval, image-derived phenotypes, shape-derived phenotypes, and HeartSSM shape descriptor as query vectors. \textbf{c}, Long-axis four-chamber view CMR images of an example subject, compared between the first and second imaging scans. \textbf{d}, Top-$K$ recall for longitudinal re-identification using random retrieval, image-derived phenotypes, shape-derived phenotypes, and HeartSSM shape descriptor as query vectors. The performance is also compared across HeartSSM shape descriptors with different number of PCs that explain different ratios (\%) of variance.}
\label{fig:retrieval}
\end{figure*}

\subsection*{Shape features for chronological age prediction}
We assess the effectiveness of shape-derived phenotypes and HeartSSM shape descriptor on chronological age prediction task \cite{shah2023age}, using a set of 9,135 healthy UK Biobank participants without reported diseases. A CatBoost regressor \cite{prokhorenkova2018catboost} is trained for age prediction using different features, including image-derived phenotypes, shape-derived phenotypes, and HeartSSM shape descriptor. The prediction performance is evaluated by the mean absolute error (MAE), Pearson correlation coefficient ($r$), and coefficient of determination ($R^2$). Fig \ref{fig:classification}b compares the performance when different features are used. It shows that shape-derived phenotypes (MAE, $5.538\pm3.946$; $r=0.378$; $R^2=0.141$) outperform image-derived phenotypes (MAE, $5.628\pm3.958$; $r=0.350$; $R^2=0.120$) with statistical significance (paired $t$-test, $p=1.0\times10^{-5}$). HeartSSM shape descriptors leads to substantial improvement of performance.

To demonstrate the advantage of the 3D+t four-chamber (4CH+t) HeartSSM model, we compare its performance with two variants, namely a 3D four-chamber (4CH) HeartSSM without motion and a 3D+t bi-ventricular (2CH+t) HeartSSM model. The 3D four-chamber HeartSSM is constructed using ED and ES frames only without intermediate frames, whereas 3D+t bi-ventricular HeartSSM is constructed without the two atria. We compare the performance of the shape descriptors of three HeartSSM models for the age prediction task. Fig.~\ref{fig:classification}b shows that the 3D+t four-chamber (4CH+t) HeartSSM (MAE, $4.249\pm3.211$; $r=0.689$; $R^2= 0.473$) significantly outperforms the 3D+t bi-ventricular (2CH+t) HeartSSM (MAE, $4.326\pm3.205$; $r=0.681$; $R^2=0.461$; paired $t$-test, $p=2.4\times10^{-5}$) and the 3D four-chamber (4CH) HeartSSM without motion information (MAE, $4.468\pm3.308$; $r=0.653$; $R^2=0.426$; paired $t$-test, $p=2.7\times10^{-28}$) for chronological age prediction. This demonstrates the effectiveness of utilising both shape and motion information of all four cardiac chambers.

\subsection*{Population-level cardiac shape retrieval}
We examine the representation ability of the HeartSSM shape descriptor for cardiac shape retrieval \cite{tangelder2004retrieval,chang2015shapenet}. Given a query HeartSSM shape descriptor, we retrieve top $K$ subjects with the most similar descriptors in the UK Biobank imaging population and assess the retrieval precision, defined as the percentage of the retrieved subjects sharing the same clinical factor as the query subject. We calculate the precision for six clinical factors, including sex, age, weight, height, BMI, and hypertension. All factors are represented as categorical variables as illustrated in Fig.~\ref{fig:retrieval}a. For comparison, we also use shape-derived phenotypes and image-derived phenotypes as query vectors for the shape retrieval task, and include random retrieval as a baseline. Fig.~\ref{fig:retrieval}b and Supplementary Table~6 report the top-$K$ precision for different clinical factors and different $K$ parameters. Each precision value is calculated using 5,000 queries. HeartSSM shape descriptor achieves the highest retrieval precision compared to shape-derived phenotypes and image-derived phenotypes for all six clinical factors.

\subsection*{Longitudinal re-identification of cardiac shapes}
Besides cardiac shape retrieval, we conduct experiments on longitudinal re-identification of cardiac four-chamber shapes. In our dataset, 4,134 participants have two CMR imaging scans with a time gap of $2.66\pm1.09$ years \cite{littlejohns2019ukbb}.  Shape-based longitudinal heart re-identification is a particularly challenging task, as the planning of the CMR scan such as the long-axis view position may vary substantially between the two scans, as illustrated in Fig.~\ref{fig:retrieval}c. Using the HeartSSM shape descriptor of a participant at the first time point, we retrieve top $K$ subjects with the most similar descriptors at the second time point and assess the re-identification accuracy. We report the top-$K$ recall in Fig.~\ref{fig:retrieval}d and Supplementary Table~7, \emph{i.e.}, the percentage of subjects that are re-identified within $K$ retrievals. For the HeartSSM shape descriptor, we use 12/17/27/58 numbers of PCs that explain 80/85/90/95\% of variance to represent the primary shape and motion patterns for longitudinal re-identification. Interestingly, we found that the reduced number of PCs results in better retrieval performance (Fig.~\ref{fig:retrieval}d, right). Using the HeartSSM shape descriptor (12 PCs, 80\% variance) as the query vector achieves $49.01\%$ recall for $K=200$, which significantly outperforms the re-identification result of image-derived phenotypes ($31.37\%$ recall@200; paired $t$-test, $p=7.1\times10^{-63}$).

\section*{Discussion}
In this study, we present learning-based methods for mesh reconstruction and shape modelling of the cardiac four chambers using a population-level imaging dataset, the UK Biobank. Our work makes five-fold contributions. First, we proposed an efficient deep learning-based pipeline for 3D+t cardiac four-chamber mesh reconstruction and motion tracking. Second, by employing the pipeline, we created a large-scale data resource of time-resolved four-chamber cardiac meshes for the UK Biobank imaging population, which comprises of nearly 100,000 participants in their mid-to-late ages. Third, we constructed HeartSSM, a population-scale shape model capturing both shape variations and motion patterns of cardiac four chambers. Fourth, we revealed the correlation of cardiac four-chamber shapes with demographics, anthropometrics, cardiovascular risk factors, and cardiac diseases. Finally, compared to conventional image-derived phenotypes, we demonstrated that shape-derived phenotypes and HeartSSM shape descriptor achieve significant improvement for cardiac disease diagnosis, age prediction, cardiac shape retrieval, and longitudinal re-identification.

The deep learning-based mesh reconstruction pipeline integrates CMR respiratory motion correction, 3D cardiac label completion via LC-U-Net \cite{xu2023complete,xu2024complete,muffoletto2024complete,ma2025cardiacflow}, and 3D+t mesh reconstruction by HeartFFDNet. The HeartFFDNet learns FFDs for joint four-chamber mesh reconstruction and motion tracking, achieving superior geometric accuracy (ASSD, $1.426\pm0.202mm$; HD90, $2.605\pm0.476mm$) as well as temporal ($0.150\pm0.016mm$) and cycle ($0.284\pm0.052mm$) consistency compared with state-of-the-art baseline methods \cite{meng2023deepmesh,gaggion2025hybridvnet}. The generalisation ability of HeartFFDNet has been verified on UK Biobank participants with a range of cardiovascular diseases. The pipeline is memory and time efficient, supporting training and inference on a single 12GB NVIDIA 3080 GPU card with 2.87 seconds of inference time. The mesh representation of the four-chamber heart provides accurate estimation of shape-derived phenotypes, in particular for the two atria. Moreover, the reconstructed meshes share the same vertex connectivity across all time frames and across all UK Biobank participants, enabling population-scale analysis of cardiac structure and function.

The statistical shape model, HeartSSM, was constructed based on the reconstructed 3D+t cardiac four-chamber meshes of 95,696 UK Biobank participants. 128 PCs of HeartSSM explained $97.45\%$ variance of the cardiac shape and motion distribution, summarising complex geometry and dynamics of the heart. HeartSSM exhibited outstanding generalisability on unseen shapes ($0.788\pm0.170mm$) as well as remarkable performance on sparse contour fitting and missing sequence completion. HeartSSM encoded a cardiac four-chamber mesh sequence into a 128-dimensional shape descriptor, serving as an informative feature vector for downstream applications.

The association studies of PCs with four-chamber shape-derived phenotypes and participant characteristics linked HeartSSM to meaningful cardiac structural and functional traits, enhancing the interpretability of the shape model. We measured vertex-wise correlation of shape variations with sex, age, BMI, hypertension, atrial fibrillation and flutter, and heart failure. Our findings revealed physiologically plausible correlations aligned with prior knowledge. More specifically, the heart size was highly associated with sex and age as expected \cite{bai2020population,burns2024genetic}. The correlation maps of BMI and hypertension displayed similar spatial distributions. Atrial fibrillation and flutter showed stronger associations with atrial shape variations than the two ventricles. Heart failure was more strongly correlated with the shape variations of the left heart, which is typically affected by the disease \cite{friedberg2014failure,uk2016failure,bozkurt2021failure}.

The representation ability and reliability of shape-derived phenotypes and HeartSSM shape descriptor were evaluated on cardiac disease classification \cite{qiao2025meshheart}, chronological age prediction \cite{shah2023age}, cardiac shape retrieval \cite{tangelder2004retrieval,chang2015shapenet}, and longitudinal re-identification tasks. Experiments validated that shape-derived phenotypes consistently outperformed image-derived phenotypes in these tasks. The HeartSSM shape descriptor, which characterises both shape and motion distributions of the hearts, provided effective features and resulted in notable improvement for all downstream applications.

However, this study has several limitations. 
First, the deep learning-based pipeline reconstructs 3D+t cardiac four-chamber meshes from segmentation maps and does not capture sub-voxel intensity patterns within CMR images. As a consequence, the quality of the reconstructed four-chamber meshes is affected by the accuracy of the input automated CMR segmentations \cite{bai2018automated}. 
Second, 3D cardiac label completion is learnt from four-chamber segmentation maps on multiple CT datasets \cite{zeng2023imagecas,wasserthal2023totalsegmentator,Zhuang2019whs,Zhuang2019mvmm,tobon2015lasc,kiricsli2013casd,metz2009ccec}, which may be different from the UK Biobank population.
Third, due to the sparsity of the CMR images, there is no available 3D+t ground truth from UK Biobank to directly evaluate the reconstructed four-chamber meshes. Therefore, we introduced surrogate metrics to evaluate the geometric accuracy of the reconstructed 3D+t meshes against sparse 2D+t contours from quality-controlled CMR segmentations \cite{bai2018automated}. 
Lastly, in the association study of vertex-wise shape variations, we only performed correlation analysis without confounders (sex, age, weight, height, BMI, etc.), as incorporating these variables via linear regression for each vertex is computationally intractable.

In summary, we have reconstructed the 3D+t four-chamber meshes of the heart for 95,695 UK Biobank participants by developing a deep learning-based pipeline. We have built a 3D+t four-chamber HeartSSM model to characterise the distribution of cardiac shape and motion. The phenotypes derived from four-chamber meshes and the HeartSSM shape descriptor demonstrate superior performance across multiple downstream tasks. The developed methods, along with the curated population-level cardiac shape and motion data resource, will benefit future research in machine learning and cardiovascular research, facilitating deeper understanding of cardiac structure and motion and offering new insights into cardiovascular diseases.

\section*{Methods}
\subsection*{UK Biobank CMR data}
In this study, we mainly use the UK Biobank CMR dataset \cite{collins2012ukbb,sudlow2015ukbb,bycroft2018ukbb,littlejohns2019ukbb} to learn the 3D+t four-chamber cardiac shape model. 102,521 participants (UK Biobank field ID 53) attended the imaging visit to one of the four imaging centres (Cheadle, Newcastle, Reading, and Bristol). After MRI safety screening and excluding participants with claustrophobia, multi-view CMR images (short-axis view, long-axis two-chamber view, long-axis four-chamber view) are available for 95,695 participants after data cleaning. Non-imaging participant characteristics, including demographics, anthropometrics, and health information, are available for 62,778 participants for this study. Among the cohort, 4,134 participants have a second CMR imaging scan, approximately 3 years after the first scan. The detailed data information is summarised in Supplementary Table~1.

\subsection*{Respiratory motion correction}
To alleviate the respiratory motion and misalignment between multi-view CMR images, we adapt an intensity-based algorithm to correct in-plane motion artefacts \cite{villard2016correction}. For short-axis view, long-axis two-chamber view, and long-axis four-chamber view images, we estimate the in-plane displacements $(\Delta x, \Delta y)$ that minimise the mean absolute intensity difference at the intersections of multi-view images. Meanwhile, we impose a Laplacian smoothness regularisation on the image intensities between consecutive short-axis view slices.

Let $I_{\mathrm{2ch}}:\mathbb{R}^2\times[1,T]\rightarrow\mathbb{R}$, $I_{\mathrm{4ch}}:\mathbb{R}^2\times[1,T]\rightarrow\mathbb{R}$ denote long-axis two-chamber and four-chamber view image sequences with $T$ time frames, and let $I^d_{\mathrm{sax}}:\mathbb{R}^2\times[1,T]\rightarrow\mathbb{R}$ denote the $d$-th slice of the short-axis view sequences with $D$ slices and $d=1,...,D$. The aim is to find in-plane displacements $(\Delta x_{\mathrm{2ch}},\Delta y_{\mathrm{2ch}})$, $(\Delta x_{\mathrm{4ch}},\Delta y_{\mathrm{4ch}})$, $(\Delta x^d_{\mathrm{sax}},\Delta y^d_{\mathrm{sax}})$ to optimise a motion correction objective. Given the displacements, the motion corrected images are formulated as:
\begin{equation}\label{eq:motion-image}
\begin{split}
\tilde{I}_{\mathrm{2ch}}(x,y,t):=I_{\mathrm{2ch}}(x+\Delta x_{\mathrm{2ch}}, y+\Delta y_{\mathrm{2ch}},t),\\
\tilde{I}_{\mathrm{4ch}}(x,y,t):=I_{\mathrm{4ch}}(x+\Delta x_{\mathrm{4ch}}, y+\Delta y_{\mathrm{4ch}},t),\\
\tilde{I}^d_{\mathrm{sax}}(x,y,t):=I^d_{\mathrm{sax}}(x+\Delta x^d_{\mathrm{4ch}}, y+\Delta y^d_{\mathrm{4ch}},t),
\end{split}
\end{equation}
and the objective function is formulated by:
\begin{equation}\label{eq:motion}
\begin{split}
\mathcal{L}&=\frac{1}{T}\sum_{t=1}^T\bigg(\sum_{(x,y)\in\mathrm{2ch\cap4ch}}
\Big|\tilde{I}_{\mathrm{2ch}}(x,y,t)-\tilde{I}_{\mathrm{4ch}}(x,y,t)\Big|\\
&+\sum_{d=1}^D\sum_{(x,y)\in\mathrm{2ch\cap (sax)}_d}
\Big|\tilde{I}^d_{\mathrm{sax}}(x,y,t)
-\tilde{I}_{\mathrm{2ch}}(x,y,t)\Big|\\
&+\sum_{d=1}^D\sum_{(x,y)\in\mathrm{4ch\cap (sax)}_d}
\Big|\tilde{I}^d_{\mathrm{sax}}(x,y,t)
-\tilde{I}_{\mathrm{4ch}}(x,y,t)\Big|\\
&+\frac{1}{2}\sum_{d=2}^{D-1}\sum_{(x,y)}\Big|\tilde{I}^{d-1}_{\mathrm{sax}}(x,y,t)+\tilde{I}^{d+1}_{\mathrm{sax}}(x,y,t)-2\tilde{I}^d_{\mathrm{sax}}(x,y,t)\Big|\bigg).
\end{split}
\end{equation}
The objective function consists of the mean absolute intensity difference at the intersections of multi-view CMR images and the Laplacian regularisation terms for each slice of the short-axis view images. We estimate the displacements by minimising objective function (\ref{eq:motion}) using Adam optimiser \cite{kingma2014adam} with a learning rate 0.1 for 1,000 epochs, and subsequently perform motion correction. The performance is evaluated using the Pearson correlation coefficient of image intensities and the Dice coefficient of segmentation maps at the intersections of multi-view CMR images.

\subsection*{LC-U-Net for 3D cardiac label completion}
We train a LC-U-Net to learn shape priors of cardiac four chambers from dense 3D CT segmentations for label completion. We collect $n=1,575$ four-chamber heart segmentation maps of CT images from multiple data sources: ImageCAS \cite{zeng2023imagecas} ($n=880$), TotalSegmentator \cite{wasserthal2023totalsegmentator} ($n=558$), WHS++ \cite{Zhuang2019whs,Zhuang2019mvmm} ($n=32$), the Left Atrium Segmentation Challenge \cite{tobon2015lasc} ($n=25$), the Coronary Artery Stenosis Detection Challenge \cite{kiricsli2013casd} ($n=48$), and the Coronary Centerline Extraction Challenge \cite{metz2009ccec} ($n=32$). For the ImageCAS dataset, the four-chamber heart segmentation maps are generated automatically by the \textit{heartchambers\_highres} model of TotalSegmentator v2 \cite{wasserthal2023totalsegmentator} following manual quality control. The four-chamber segmentations of all CT images follow a consistent segmentation protocol, including LV, LVM, RV, LA, and RA label classes. All 3D CT segmentations are rescaled and rigidly transformed to an atlas space \cite{Zhuang2016atlas} at $2mm$ resolution, cropped to a fixed size $(L,W,H)$ with $L=W=96$ and $H=128$. The 1,575 CT segmentations are divided into subsets of 945/157/473 data for training/validation/test. Note that the original CT images are not needed for cardiac label completion.

As depicted in Supplementary Fig.~1, we simulate sparse multi-view segmentations based on the dense 3D four-chamber segmentations following the settings in previous work \cite{xu2023complete,xu2024complete,muffoletto2024complete,ma2025cardiacflow}. To address motion artefacts that are not fully resolved in the motion correction step, we introduce random in-plane displacements to the simulated short-axis view segmentations, following a Gaussian distribution $\Delta x,\Delta y\sim\mathcal{N}(0,\sigma)$ with a standard deviation $\sigma=2~mm$ . Both dense 3D and sparse multi-view segmentations are scaled by a random factor drawn from a uniform distribution $\mathcal{U}([0.8,1.2])$. 
Using the simulated sparse multi-view segmentations as input, the LC-U-Net is trained to predict the dense 3D four-chamber segmentations for label completion. The mean squared error loss between the completed and original four-chamber segmentation maps is minimised  using the Adam optimiser with a learning rate of $10^{-4}$ and a batch size of $1$ for 200 epochs. The dropout rate for each layer of the LC-U-Net is set to $0.2$ during training. Once trained, LC-U-Net is applied to complete the sparse 2D+t multi-view CMR segmentations into 3D+t dense cardiac four-chambers segmentations for 95,695 UK Biobank participants.

\subsection*{Four-chamber heart template}
Based on the four-chamber heart shapes from 3D CT segmentations \cite{zeng2023imagecas,wasserthal2023totalsegmentator,Zhuang2019whs,Zhuang2019mvmm,tobon2015lasc,kiricsli2013casd,metz2009ccec}, we construct a four-chamber heart template that consists of a template segmentation and a template mesh. First, the four-chamber segmentation template is created by averaging 945 CT segmentation maps in the training set, which have been rescaled and rigidly aligned to an atlas space \cite{Zhuang2016atlas}. Then, we compute the convex hull meshes for five cardiac structures: LV endocardium (LV-endo), LV endocardium (LV-epi), RV, LA, and RA. These convex hull meshes are smoothed and remeshed as initial surfaces for a deformable model \cite{schuh2017deformable}, which deforms the convex hull meshes (LV-endo, LV-epi, RV, LA, RA) along surface normal directions to match the template segmentation, guided by the signed distance function derived from the template segmentation. The LV-epi surface is produced by inflating the LV-endo surface, so that the LV-epi and LV-endo meshes have corresponding vertices with the same connectivity, facilitating the measurement of LV myocardial wall thickness. The template mesh has 27,034 vertices in total, with 6,141/6,141/5,696/4,305/4,751 vertices for LV-endo/LV-epi/RV/LA/RA respectively. Supplementary Fig.~2 visualises each structure of the cardiac four-chamber template mesh.

\subsection*{HeartFFDNet for dynamic mesh reconstruction}
As illustrated in Fig.~\ref{fig:pipeline}b, HeartFFDNet learns FFDs \cite{sederberg1986ffd} from input 3D+t dense segmentations of the four chambers and deforms a template mesh into 3D+t four-chamber meshes for an individual heart. Given 3D+t four-chamber segmentations of size $(T,L,W,H)$ and a template segmentation as input, HeartFFDNet employs a U-Net architecture \cite{ronneberger2015unet} to predict B-spline FFDs at two scales, which have control grids of sizes $(L/16,W/16,H/16,3)$ and $(L/8,W/8,H/8,3)$. The four-chamber template mesh is warped by predicted multi-scale FFDs into an initial mesh for the following mesh deformation and motion tracking. Then, HeartFFDNet predicts FFDs of size $(L/4,W/4,H/4,3)$ for each of the $T$ time frames, which deform the initial mesh into a 3D+t dynamic sequence of four-chamber meshes.

To train HeartFFDNet, we formulate a loss function to minimise the distance between the deformed 3D+t four-chamber meshes predicted by HeartFFDNet and the pseudo ground-truth 3D+t point clouds, which are derived from 3D+t segmentations using the marching cubes algorithm \cite{lorensen1998marching}. We denote the deformed 3D+t four-chamber meshes as 
\begin{equation}\label{eq:mesh}
\mathcal{M}_t=\{\mathcal{M}_t^c\}_{c\in C},~~C=\{\mathrm{LV_{endo},~LV_{epi},~RV,~LA,~RA}\},
\end{equation}
where $\mathcal{M}_t^c$  denotes the mesh for structure $c$ at time frame $t=1,...,T$. Each mesh is represented as $\mathcal{M}_t^c=(\mathcal{V}_t^c, \mathcal{E}^c, \mathcal{F}^c)$, where $\mathcal{V},\mathcal{E},\mathcal{F}$ define the sets of vertices, edges and faces. Since the mesh of each structure has the same connectivity across all time frames, the edge set $\mathcal{E}^c$, the face set $\mathcal{F}^c$, and the number of vertices $|\mathcal{V}^c|$ do not depend on the frame $t$.

The loss function consists of a reconstruction loss $\mathcal{L}_{\mathrm{recon}}$, an edge loss $\mathcal{L}_{\mathrm{edge}}$, a curvature loss $\mathcal{L}_{\mathrm{curv}}$, a temporal consistency loss $\mathcal{L}_{\mathrm{temp}}$, and a cycle consistency loss $\mathcal{L}_{\mathrm{cycle}}$. Given the pseudo ground-truth 3D+t point clouds $\mathcal{P}_t^c\subset\mathbb{R}^3$ for each structure $c\in C$, the reconstruction loss $\mathcal{L}_{\mathrm{recon}}$ is defined by the ASSD between predicted 3D+t four-chamber meshes and point clouds:
\begin{equation}\label{eq:loss-recon}
\begin{split}
&\mathcal{L}_{\mathrm{recon}}=\\
&\frac{1}{T}\sum_{t=1}^T\sum_{c\in C}\Big(
\frac{1}{|\mathcal{V}_t^c|}\sum_{v\in\mathcal{V}_t^c}\min_{p\in\mathcal{P}_t^c}\|v-p\|
+\frac{1}{|\mathcal{P}_t^c|}\sum_{p\in\mathcal{P}_t^c}\min_{v\in\mathcal{V}_t^c}\|v-p\|
\Big)
\end{split}
\end{equation}
To regularise the smoothness of the meshes, we use an edge loss to minimise the standard deviation of edge length:
\begin{equation}\label{eq:loss-edge}
\mathcal{L}_{\mathrm{edge}}=\frac{1}{T}\sum_{t=1}^T\sum_{c\in C}
\bigg(\frac{1}{|\mathcal{E}^c|}\sum_{(i,j)\in\mathcal{E}^c}\left(e_{t,i,j}^c-\bar{e}_t^c\right)^2\bigg)^{\frac{1}{2}}
\end{equation}
where $e_{t,i,j}^c:=\|v_{t,i}^c-v_{t,j}^c\|$ is the edge length between the vertex $i$ and $j$, and $\bar{e}^c_t:=\frac{1}{|\mathcal{E}^c|}\sum_{(i,j)\in\mathcal{E}^c}e_{t,i,j}^c$ is the mean edge length for structure $c$ and frame $t$. 

The curvature loss $\mathcal{L}_{\mathrm{curv}}$ encourages all meshes to have the identical curvature distribution as the template mesh, enforcing the shape correspondence across all subjects and time frames.  For the $i$-th vertex $v_{t,i}^c\in\mathcal{V}_t^c$ of structure $c$ at frame $t$, its mean curvature $H_{t,i}^c\in\mathbb{R}$ is estimated by \cite{nealen2006laplacian,do2016differential},
\begin{equation}\label{eq:curvature}
H_{t,i}^c = -\frac{1}{2}\Delta v_{t,i}^c\cdot n_{t,i}^c\approx-\frac{1}{2} (L\mathbf{v}_{t}^c)_i\cdot n_{t,i}^c,
\end{equation}
where $n_{t,i}^c\in\mathbb{R}^3$ is the unit normal vector at vertex $v_{t,i}^c$, $\Delta$ is the Laplacian-Beltrami operator \cite{nealen2006laplacian,do2016differential}, $L\in\mathbb{R}^{|\mathcal{V}^c|\times|\mathcal{V}^c|}$ is the graph Laplacian matrix, and $\mathbf{v}_{t}^c\in\mathbb{R}^{|\mathcal{V}^c|\times 3}$ is the vectorised representation of all vertices $v_{t,i}^c$ of the structure $c$ at time frame $t$. Let $\mathbf{H}_t^c\in\mathbb{R}^{|\mathcal{V}^c|}$ denote the curvatures of all vertices of structure $c$ at frame $t$, and $\mathbf{H}_0^c\in\mathbb{R}^{|\mathcal{V}^c|}$ denote the curvatures of the template mesh. The curvature loss $\mathcal{L}_{\mathrm{curv}}$ is defined by,
\begin{equation}\label{eq:loss-curv}
\mathcal{L}_{\mathrm{curv}}=\frac{1}{T}\sum_{t=1}^{T}\sum_{c\in C}1-r(\mathbf{H}_t^c,\mathbf{H}_0^c),
\end{equation}
where $r(\mathbf{H}_t^c,\mathbf{H}_0^c)\in[-1,1]$ is the Pearson correlation coefficient between the curvatures of the reconstructed meshes and the template mesh.

The temporal consistency loss  $\mathcal{L}_{\mathrm{temp}}$ regularises the deformation between two consecutive frames, which is defined as,
\begin{equation}\label{eq:loss-temp}
\mathcal{L}_{\mathrm{temp}}=\frac{1}{T-1}\sum_{t=1}^{T-1}\sum_{c\in C}\frac{1}{|\mathcal{V}^c|}\sum_{i=1}^{|\mathcal{V}^c|}\|v_{t+1,i}^c-v_{t,i}^c\|.
\end{equation}
The cycle consistency loss $\mathcal{L}_{\mathrm{cycle}}$ is formulated as,
\begin{equation}\label{eq:loss-cycle}
\mathcal{L}_{\mathrm{cycle}}=\sum_{c\in C}\frac{1}{|\mathcal{V}^c|}\sum_{i=1}^{|\mathcal{V}^c|}\|v_{T,i}^c-v_{1,i}^c\|,
\end{equation}
which minimises the distance between the meshes at the starting ($t=1$) and end ($t=T$) frame of a cardiac cycle, encouraging the heart to return to its original shape after one complete cycle \cite{ma2025cardiacflow}. The final loss function is a combination of the loss terms defined as:
\begin{equation}\label{eq:loss-all}
\begin{split}
\mathcal{L}_{\mathrm{all}}=\mathcal{L}_{\mathrm{recon}}
&+\lambda_{\mathrm{edge}} \cdot \mathcal{L}_{\mathrm{edge}}+\lambda_{\mathrm{curv}} \cdot \mathcal{L}_{\mathrm{curv}}\\
&+\lambda_{\mathrm{temp}} \cdot \mathcal{L}_{\mathrm{temp}}+\lambda_{\mathrm{cycle}} \cdot \mathcal{L}_{\mathrm{cycle}},
\end{split}
\end{equation}
where the weights $\lambda_{\mathrm{edge}}=0.5$, $\lambda_{\mathrm{curv}}=1.0$, $\lambda_{\mathrm{temp}}=0.1$, and  $\lambda_{\mathrm{cycle}}=0.2$ are empirically chosen. The loss function is minimised using the Adam optimiser with a learning rate $10^{-4}$ and a batch size of 1 for 200 epochs, using 600 training and 100 validation subjects from the UK Biobank dataset. The performance of HeartFFDNet is evaluated using 1,000 test subjects with multi-view 2D+t four-chamber contours derived from quality-controlled CMR segmentations. Please see Supplementary Table~2 for detailed data information.

\subsection*{Learning the HeartSSM shape model}
We learn a 3D+t cardiac four-chamber statistical shape model, named as HeartSSM, from the UK Biobank imaging population using incremental PCA \cite{weng2003ipca,zhao2006ipca,brand2006svd}. In order to perform PCA, the 3D+t cardiac four-chamber meshes reconstructed by the deep learning-based pipeline are rigidly aligned to the space of the template mesh. Then, we vectorise all $|V|=27,034$ vertices of each 3D+t four-chamber mesh sequence with $T=50$ time frames into a shape vector $\mathbf{v}\in\mathbb{R}^{3T|V|}$. 
Due to memory limitation, it is computationally intractable to apply standard PCA \cite{abdi2010pca,gewers2021pca,greenacre2022pca} to mesh sequences from a large population. Given a training set with $N$ subjects, standard PCA explicitly constructs a covariance matrix for $N$ vectorised shapes $\mathbf{v}$ of dimensionality $3T|V|$, which amounts to a quadratic memory cost of $O(N^2T^2|V|^2)$. To address the memory limitation, HeartSSM is constructed using incremental PCA \cite{weng2003ipca,zhao2006ipca,brand2006svd} for $N=64,000$ training subjects (see Supplementary Table~2), which updates the mean shape and the PCs progressively using mini-batches. We employ a batch size of 128 for incremental PCA and set the number of PCs to $M=128$.

After training, HeartSSM approximates each 3D+t four-chamber shape vector $\mathbf{v}$ by,
\begin{equation}\label{eq:ssm}
\mathbf{v}\approx\bar{\mathbf{v}}+\mathbf{P}\mathbf{w},
\end{equation}
where $\bar{\mathbf{v}}\in\mathbb{R}^{3T|V|}$ is the mean shape vector capturing both shape and motion of the four chambers, $\mathbf{P}\in\mathbb{R}^{3T|V|\times M}$ denotes $M=128$ orthogonal PCs learnt from the incremental PCA, and $\mathbf{w}\in\mathbb{R}^{M}$ contains the weights of the PCs. Consequently, the cardiac shape variation and motion pattern for each of the 95,695 UK Biobank participants can be represented by the low-dimensional vector $\mathbf{w}$, which is also called the \textit{HeartSSM shape descriptor}. For any unseen test shape vector $\mathbf{v}$, the HeartSSM shape descriptor $\mathbf{w}$ can be estimated by,
\begin{equation}\label{eq:ssm-fit}
\mathbf{w}=\mathbf{P}^{\top}(\mathbf{v}-\bar{\mathbf{v}}).
\end{equation}
The generalisation ability of HeartSSM is evaluated by fitting 10,000 unseen test four-chamber mesh sequences (Fig.~\ref{fig:heartssm}d).

HeartSSM can fit onto 3D+t cardiac meshes with missing spatial or temporal information, such as spare contour fitting (Supplementary Fig.~3) or missing sequence completion (Supplementary Fig.~4). Given incomplete meshes, we optimise the HeartSSM shape descriptor $\mathbf{w}$ to minimise the distance between the reconstructed meshes and incomplete meshes, which can be either sparse contours or a mesh sequence with missing time frames. We use the Adam optimiser with 0.05 learning rate (500 iterations) for sparse contour fitting and 0.1 learning rate (200 iterations) for missing sequence completion. The fitting ability of HeartSSM is evaluated using 1,000 test subjects with manual quality control. Please see Supplementary Table~2 for detailed test data information.

\subsection*{Correlation analysis of shape variation}
For the cardiac four-chamber mesh of each subject, we assess vertex-wise shape variation by comparing the individual subject mesh to the population mean mesh derived from HeartSSM. For each vertex $v_t\in\mathbb{R}^3$ at frame $t$, the shape variation is quantified by the signed distance $s_t\in\mathbb{R}$ from the vertex $v_t$ to the corresponding vertex $\bar{v}_t$ on the mean mesh, along the normal direction $\bar{n}_t$ at the vertex $\bar{v}_t$, \emph{i.e.},
\begin{equation}\label{eq:shape-variation}
    s_t:=\langle v_t-\bar{v}_t, \bar{n}_t\rangle.
\end{equation}
A positive shape variation ($s_t>0$) indicates that the individual four-chamber mesh expands outwards along the surface normal direction relative to the mean mesh. To summarise the shape variation across a cardiac cycle, we compute the mean shape variation $s=\frac{1}{T}\sum_{t=1}^T s_t$ at each vertex across all time frames. Subsequently, we perform correlation analysis of vertex-wise shape variation against each attribute $y$ (sex, age, BMI, hypertension, atrial fibrillation and flutter, and heart failure). For each vertex, we compute the Pearson correlation coefficient between the shape variation $s$ and the attributes $y$ for 62,778 UK Biobank participants.

\subsection*{Cardiovascular disease classification}
We perform classification for six types of cardiovascular diseases: hypertension ($n=13,453$), angina pectoris ($n=2,606$), acute myocardial infraction ($n=1,395$), chronic ischaemic heart disease ($n=3,940$), atrial fibrillation and flutter ($n=2,730$), and heart failure ($n=804$). Please see Supplementary Table~1 for detailed data information. We compare classification performance across different features: confounders (age, sex, weight, height, BMI), image-derived phenotypes with/without confounders, shape-derived phenotypes with/without confounders, and HeartSSM shape descriptor with/without confounders. We include 17 image-derived or shape-derived phenotypes: left ventricular myocardial mass (LVM), ventricular ED and ES volumes (LVEDV, LVESV, RVEDV, RVESV), atrial maximum and minimum volumes (LAMAXV, LVMINV, RAMAXV, RAMINV), four-chamber stroke volumes (LVSV, RVSV, LASV, RASV), and four-chamber ejection fractions (LVEF, RVEF, LAEF, RAEF). We employ two classifiers, logistic regression and XGBoost \cite{chen2016xgboost}, with default hyperparameters except that the learning rate of XGBoost is set to 0.05. The classifiers are trained using four-fold cross validation on 62,778 UK Biobank participants with available demographic and disease information.

\subsection*{Chronological age prediction}
For chronological age prediction \cite{shah2023age}, we train a CatBoost model \cite{prokhorenkova2018catboost} for age regression using four-fold cross validation on 9,135 healthy UK Biobank participants, with 0.05 learning rate and 1,000 iterations. We compare different features, including image-derived phenotypes, shape-derived phenotypes, and HeartSSM shape descriptors. Apart from the proposed HeartSSM constructed using 3D+t four-chamber meshes, called HeartSSM (4CH+t), we also construct two variants of HeartSSM, using 3D four-chamber meshes of ED and ES frames only without motion, called HeartSSM (4CH), and using 3D+t bi-ventricular meshes without two atria, called HeartSSM (2CH+t), respectively.

\subsection*{Cardiac shape retrieval}
In the shape retrieval task, each of the UK Biobank participants is represented by a shape-based feature vector $x$, which can be image-derived phenotypes, shape-derived phenotypes, or HeartSSM shape descriptor. Each participant is assigned to a group $y$ according to sex, age, weight, height, BMI, or hypertension disease (Fig.~\ref{fig:retrieval}a). We randomly select $M=5,000$ query subjects $(x^q,y^q)$. The aim of shape retrieval is to retrieve a subject $(x^r,y^r)$ from the UK Biobank population, based on the query vector $x^q$, and assess whether the retrieved subject belongs to the same group as the query subject, \emph{i.e.}, $y^r=y^q$. For each query vector $x_i^q$, $i=1,...,M$, we run KNN algorithm to retrieve $K$ nearest neighbours $x_{i_k}^r$, $k=1,...,K$, from all participants except for the query subject. Then, the top-$K$ precision is measured by
\begin{equation}\label{eq:precision}
\mathrm{Precision@K}=\frac{1}{M}\sum_{i=1}^M\frac{1}{K}\sum_{k=1}^K\mathds{1}_{\{y_{i_k}^r=y_i^q\}}\times100\%,
\end{equation}
where the indicator function $\mathds{1}_{\{a=b\}}=1$ if and only if $a=b$.

\subsection*{Shape-based longitudinal re-identification}
In the longitudinal re-identification task, we utilise a subset of the UK Biobank population, which contains 4,134 participants with two CMR imaging scans $2.66\pm1.09$ years apart. We use the query vector $x_i^{(2)}$ of the second CMR scan to re-identify the $i$-th subject from the population, based on the feature vector $x_i^{(1)}$ derived from the first CMR scan. We employ KNN algorithm to retrieve $K$ subjects $x_{i_k}^{(1)}$, and the top-$K$ recall is defined as
\begin{equation}\label{eq:precision}
\mathrm{Recall@K}=\frac{1}{M}\sum_{i=1}^M\sum_{k=1}^K\mathds{1}_{\{i_k=i\}}\times100\%,
\end{equation}
where $i_k=i$ means the subject is successfully re-identified within $K$ retrievals.

\section*{Data availability}
Raw imaging data and non-imaging participant characteristics are available from UK Biobank to approved researchers via a standard application process at \href{https://www.ukbiobank.ac.uk/register-apply}{https://www.ukbiobank.ac.uk/register-apply}.

\section*{Code availability}
The code for this study is publicly available at the Github repository: \href{https://github.com/m-qiang/HeartSSM}{https://github.com/m-qiang/HeartSSM}.

\bibliography{ref}

\section*{Acknowledgements}
This research was conducted using the UK Biobank Resource under Application Number 18545. We thank all UK Biobank participants and staff.
This work is supported by the EPSRC grants (EP/W01842X/1, EP/Z531297/1) and the BHF New Horizons Grant (NH/F/23/70013). 
S.N. is supported by the National Institutes of Health (R01-HL152256), the European Research Council (PREDICT-HF 864055), the British Heart Foundation (RG/20/4/34803), the EPSRC (EP/X012603/1 and EP/P01268X/1), the Technology Missions Fund under the EPSRC (EP/X03870X/1), and the Alan Turing Institute.
D.P.O. is supported by the Medical Research Council (MC\_UP\_1605/13), the NIHR Imperial College Biomedical Research Centre, and the British Heart Foundation (RG/F/24/110138, RE/24/130023, CH/F/24/90015).
P.M.M. acknowledges generous personal support from the Edmond J. Safra Foundation and Lily Safra, an NIHR Senior Investigator Award, Rosalind Franklin Institute, and the UK Dementia Research Institute, which is funded predominantly by the UKRI Medical Research Council.

\section*{Competing interests}
S.N. has received research funding from GSK, ANSYS and Synopsis. D.P.O. has consulted for Bayer AG and Bristol-Myers-Squibb, and received grants from Bayer and Calico Labs. P.M.M. has received consultancy or speaker fees from Roche, Merck, Biogen, Rejuveron, Sangamo, Nodthera and Novartis. P.M.M. has received research or educational funds from Biogen, Novartis, Merck and GlaxoSmithKline. The remaining authors declare no competing interests.

\end{document}